\documentstyle[aps,epsf]{revtex}
\begin{document}
\draft
\title{Selectivity of the  
$^{16}$O(e,e$'$pp) reaction to discrete final states}
\author{C.~Giusti and F.~D.~Pacati}
\address{Dipartimento di Fisica Nucleare e Teorica dell'Universit\`a, Pavia\\
and Istituto Nazionale di Fisica Nucleare, Sezione di Pavia, Italy}
\author{K.~Allaart and W.~J.~W.~Geurts}%
\address{Department of Physics and Astronomy, Vrije Universiteit,\\
      De Boelelaan 1081, 1081 HV Amsterdam, The Netherlands}
\author{W.~H.~Dickhoff}
\address{Department of Physics, Washington University,
         St.Louis, MO 63130, USA}
\author{H.~M\"uther}
\address{
Institut f\"ur Theoretische Physik, Universit\"at T\"ubingen,\\
Auf der Morgenstelle 14, D-72076 T\"ubingen, Germany}
\date{\today}
\maketitle

\begin{abstract}%
Resolution of discrete final states in the $^{16}$O(e,e$'$pp)$^{14}$C
reaction may provide an interesting tool to discriminate between
contributions from one- and two-body currents in this reaction.
This is based on the observation that the $0^+$ ground state and first $2^+$
state of $^{14}$C are reached predominantly by the removal of a $^1S_0$
pair from $^{16}$O in this reaction, whereas other states mostly arise by the 
removal of a $^3P$ pair. This theoretical prediction has been
supported recently by an analysis of the pair momentum distribution
of the experimental data\cite{OnAl}. In this paper we present results of
reaction calculations performed
in a direct knock-out framework where final-state interaction and 
one- and  two-body currents are included.
The two-nucleon overlap integrals 
are obtained from a calculation of the two-proton spectral function of 
$^{16}$O and include both long-range and short-range correlations.
The kinematics 
chosen in the calculations is relevant for recent experiments at
NIKHEF and Mainz.
We find that 
the knock-out of a $^3P$ proton pair is largely due to the
(two-body) $\Delta$-current.
The $^1S_0$ pair knock-out, on the other hand, 
is dominated by contributions from the one-body current and 
therefore sensitive to two-body 
short-range correlations. 
This opens up good perspectives for the study of these correlations in
the $^{16}$O(e,e$'$pp) reaction involving the lowest few states in $^{14}$C.
In particular the longitudinal structure function $f_{00}$, 
which might be separated with super-parallel kinematics, turns out to be quite 
sensitive to the NN potential that is adopted in the calculations.
\end{abstract}

\section{Introduction}
Exclusive (e,e$'$pp) reactions on nuclei have recently been added to the
rich set of tools exploring the nucleus with the electromagnetic 
interaction\cite{OnAl}.
It is expected that this new tool may finally be able to clarify the nature
and influence of short-range correlations (SRC) in low-energy nuclear 
phenomena.
Several early theoretical papers~\cite{Go,Kr} established a link between
two-nucleon removal cross sections and the two-nucleon density
matrix~\cite{Go} or the two-nucleon spectral function~\cite{Kr} which
contain information related to SRC.
A somewhat different perspective to this issue has been explored in Refs.\
\cite{CzGo,deF}.
The anticipated availability of this reaction generated renewed
theoretical interest~\cite{Cio,Fa,Bo} in the reaction description and
the calculation of the two-nucleon spectral function.
In a series of publications~\cite{GP,GPR,Ox,GP97,GPex}
a reaction description of electron- and photon-induced two-nucleon
emission processes has been established.
Indeed, it appears from these studies
that the most promising reaction to study short-range
phenomena involves the
(e,e$'$pp) channel, where the effect of meson-exchange currents and 
$\Delta$-isobars is less dominant as compared to the (e,e$'$pn) 
and ($\gamma$,NN) processes.

In its most recent form~\cite{GP97} the description of the (e,e$'$pp)
excitation process includes the contribution of the usual one-body terms
as well as those
two-body currents which involve the intermediate excitation of the
$\Delta$-isobar.
The de-excitation of the $\Delta$ after absorption of the photon
or the excitation of the $\Delta$ before absorption of the photon proceeds
by exchange of a pion with another nucleon.
In the present work an improvement of the dynamic aspects of the propagation
of the $\Delta$-isobar is taken into account~\cite{WAGP}.
A treatment of $\Delta$-propagation involving the exchange of rho-mesons is
not included at present.
The treatment of the final state interaction of the outgoing protons with
the remaining nucleus is treated by neglecting their mutual interaction but
including the distorting effect of their interaction with the remaining
nucleons in terms of an optical potential.
The latter distortion of the individual protons is constrained by
experimental data obtained from elastic scattering of nucleons off nuclei.
The approximation to neglect the interaction between the two outgoing
protons has been justified in the past by arguing that the pair of protons
will leave the nucleus largely back to back making this type of final state
interaction less important.
This issue should be further studied in the future since there is no a
priori dominance of the effects of correlations before or after the
absorption of the photon as emphasized in Ref.\ \cite{deF}.
It is, however, possible that angular momentum and parity restrictions
associated with the transition to specific discrete final states in the
remaining nucleus, may filter the importance of this type of final state
interaction.
The possibility to analyze different final states in the reaction has already
been explored in Ref.\ \cite{GP97}. The separation of some of the low-lying
final states has recently been realized experimentally
at the NIKHEF~\cite{OnAl} and Mainz~\cite{Rosner} facilities.
As will be shown in this paper, this feature plays a crucial role in
clarifying the distinction between transitions that are strongly influenced
by SRC and those where two-body transition currents
play a dominant role.
A related description of the reaction process has been presented in
Refs.\ \cite{Ry1,Ry2}.

The critical information about SRC in the transition
to the final A-2 state is incorporated in the two-body spectral function at
the corresponding energy. 
At low missing energy, it represents the probability density for the removal
of a pair of nucleons (protons in the present work) from the $^{16}$O ground
state to a specific discrete final state in $^{14}$C.
Since this removal amplitude involves nucleons close to the Fermi energy,
the accurate description of this process requires a careful treatment of
the influence of low-energy, or long-range, correlations associated with
the soft-surface features of the $^{16}$O nucleus.
The latter feature has not been included in Ref.~ \cite{GP97}, but 
is incorporated in Ref.\ \cite{GeAl2}.
It is the purpose of the present work to combine the reaction description
of the two-proton removal process of Ref.\ \cite{GP97}
with the many-body calculation of the two-particle spectral function
in $^{16}$O\ of Ref.\ \cite{GeAl2} in order
to calculate cross sections for the triple-coincidence experiments
performed at NIKHEF and Mainz.

The calculation of the two-body spectral function in Ref.\ \cite{GeAl2}
includes the dressing of individual nucleons through their coupling to
low-lying core excitations. In addition, the reduced presence of these
nucleons at low energy associated with strength removal due to the
influence of SRC is incorporated~\cite{GeAl1}. This yields theoretical
spectroscopic factors for low-lying states in $^{15}$N which represent the
closest agreement with experiment~\cite{Leu94} to date.
Consistency between the two aspects of the calculations (long-range vs.
short-range) is ensured by employing the same effective interaction
(G-matrix) in the calculation of the long-range correlations which
is responsible for the removal of single-particle strength.
Although the appearance of high-momentum nucleons in the ground state is
implied by SRC, their presence is only apparent at high
excitation energy in the A-1 system~\cite{MD,MPD}.
The corresponding cross section for the removal of high-momentum protons
from $^{16}$O in the (e,e$'$p) reaction has recently been
calculated in Ref.\ \cite{PoRa}.
Although these cross sections are large enough to be detectable
at these high energies, other competing processes will also be present
making a clear-cut identification of SRC in the (e,e$'$p) reaction
difficult.

This elusive consequence of SRC in the (e,e$'$p) reaction does not pertain
to the removal of a pair of nucleons leading to a discrete final state
in the A-2 system since few other competing processes are present.
The strongly reduced probability for a pair of protons to be in close
proximity will unavoidably lead to the presence of high-momentum components
in their relative momentum wave function.
The character and strength of these high-momentum components depends
on certain aspects of short-range phenomena which are described
differently by different nucleon-nucleon (NN) interactions.
Sensitivity to the choice of the NN interaction in describing
pairs with high relative momentum in the two-body spectral
function has been established in Ref.~\cite{GeAl2}.
It is hoped that a realistic treatment of the reaction process combined
with a detailed many-body treatment of the spectral function in conjunction
with new experimental data will produce the first clear and unambiguous
determination of SRC in nuclei.

As will be discussed in this work, the presence of several different
discrete final states in the reaction allows the calibration of the
contribution of one- and two-body current contributions and makes it
possible to assess which transitions are most sensitive to the
presence of SRC with some confidence.
This feature makes the $^{16}$O nucleus a prime target for this analysis,
unlike the $^4$He nucleus which does not yield any bound states upon 
the removal of two protons.
In Sec. II of this paper the essential ingredients of the description of
the (e,e$'$pp) reaction and the calculation of the two-particle spectral
function are summarized.
The results are discussed in Sec. III, while conclusions are drawn in Sec.
IV.


\section{Calculation of the (\lowercase{e,e$'$pp}) cross 
section}

\subsection{Reaction mechanism}

The triple coincidence cross section for the reaction induced by an electron, 
with momentum ${\mbox{\boldmath $p$}}_0$, where two nucleons, with momenta
${\mbox{\boldmath $p$}}'_1$ and ${\mbox{\boldmath $p$}}'_2$,
are ejected from a nucleus is given, in the one-photon exchange 
approximation, by the contraction between a lepton and a hadron tensor. If the 
effect of the nuclear Coulomb field on the incident and the outgoing electrons 
is neglected, the Lorentz condition for the M\"{o}ller potential and the 
continuity equation 
for the hadronic current make it possible to separate the longitudinal and 
transverse components of the interaction and to write the cross section as a 
linear combination of independent structure functions. For an unpolarized 
electron, after integration over the energy of one of the emitted nucleons 
($E'_2$), the cross section is expressed in terms of 
six structure functions as~\cite{GP,GPR,Ox},

\begin{eqnarray}
\frac{{\rm d}^{8}\sigma}{{\rm d}p'_{0}{\rm d}\Omega'_{0}
{\rm d}E'_{1}{\rm d}
\Omega'_{1}{\rm d}\Omega'_{2}} = \frac{\pi e^2}{2 q} 
\Gamma_{\rm V} \, 
\Omega_{\rm f} f_{\rm{rec}} &[&2\epsilon_{\rm L}f_{00}+f_{11}-
\epsilon\,(f_{1-1} \cos2\alpha+\bar{f}_{1-1} \sin2\alpha)   \nonumber \\
   \nonumber \\
    & + & \sqrt{\epsilon_{\rm L}(1+\epsilon)}(f_{01}\cos\alpha+\bar{f}_{01}
\sin\alpha)], \label{eq:cs}
\end{eqnarray}
where $e^2/4\pi \simeq 1/137$, ${\mbox{\boldmath $p$}}'_0$ is the momentum of
the scattered electron, $\alpha$ is the angle 
between the plane of the electrons and the plane containing the momentum 
transfer ${\mbox{\boldmath $q$}}$ and ${\mbox{\boldmath $p$}}'_1$.
The quantity 

\begin{equation}
\epsilon = \left(1-\frac{2{\mbox{\boldmath $q$}}^{2}}{q^{2}_{\mu}}\tan^{2}
\frac{\theta}{2}\right)^{-1}
\end{equation}
measures the polarization of the virtual photon exchanged by the electron
scattered at an angle $\theta$ and 
\begin{equation}
\epsilon_{\rm L} = -\frac{q^{2}_{\mu}}{{\mbox{\boldmath $q$}}^{2}}\epsilon,
\end{equation}
where $q^{2}_{\mu}=\omega^{2}-{\mbox{\boldmath $q$}}^{2}$, with 
$\omega=p_0-p'_0$ and 
${\mbox{\boldmath $q$}}={\mbox{\boldmath $p$}}_0-{\mbox{\boldmath $p$}}'_0$, is the four-momentum transfer. The factor
\begin{equation}
\Gamma_{\rm V} = \frac{e^2}{8\pi^3} \, \frac{p'_0}{p_0} \, 
\frac{q}{q^{2}_{\mu}} \, \frac{1}{\epsilon-1},
\end{equation}
is the flux of virtual photons, $\Omega_{\rm f} =p'_{1}E'_{1}p'_{2}E'_{2}$ 
is the phase-space factor and 
\begin{equation}
f_{\rm rec}^{-1} = 1- \frac{E'_2}{E_{\rm B}} 
\frac{{\mbox{\boldmath $p$}}'_{2} \cdot {\mbox{\boldmath $p$}}_{\rm B}}
{{{\mbox{\boldmath $p$}}'_{2}}^{2}}
\end{equation}
is the inverse of the recoil factor.
The quantity $E_{\rm B}$ is the total relativistic 
energy of the residual nucleus with momentum 
${\mbox{\boldmath $p$}}_{\rm B}={\mbox{\boldmath $q$}}-
{\mbox{\boldmath $p$}}'_{1}-{\mbox{\boldmath $p$}}'_{2}$.

The structure functions $f_{\lambda \lambda'}$ represent the response of the 
nucleus to the longitudinal ($\lambda = 0$) and transverse ($\lambda = \pm 1$) 
components of 
the electromagnetic interaction and only depend on  $\omega$, $q$, $p'_{1}$ 
$p'_{2}$ and the angles $\gamma_{1}$ between ${\mbox{\boldmath $q$}}$ and
${\mbox{\boldmath $p$}}'_{1}$, 
 $\gamma_{2}$ between ${\mbox{\boldmath $q$}}$ and 
 ${\mbox{\boldmath $p$}}'_{2}$ and $\gamma_{12}$ between
 ${\mbox{\boldmath $p$}}'_{1}$ 
and ${\mbox{\boldmath $p$}}'_{2}$~\cite{GP}.
They result from suitable combinations of the 
components of the hadron tensor~\cite{GP,Ox} and are thus given by 
bilinear combinations of the Fourier transforms of the transition matrix 
elements of the nuclear charge-current density operator taken between initial 
and final nuclear states
\begin{equation}
J^{\mu}({\mbox{\boldmath $q$}}) = \int \langle\Psi_{\rm{f}}
|\hat{J}^{\mu}({\mbox{\boldmath $r$}})|\Psi_{\rm{i}}\rangle
{\rm{e}}^{\,{\rm{i}}
{\footnotesize {\mbox{\boldmath $q$}}}
\cdot
{\footnotesize {\mbox{\boldmath $r$}}}
} {\rm d}{\mbox{\boldmath $r$}} .     \label{eq:jm}
\end{equation}
These integrals represent the basic ingredients of the calculation.

If the residual nucleus is left in a discrete eigenstate of its Hamiltonian,
i.e. for an exclusive process, and under the assumption of a direct knock-out
mechanism, the matrix elements of Eq.~(\ref{eq:jm}) can be written
as~\cite{GP,GP97}
\begin{equation}
J^{\mu}({\mbox{\boldmath $q$}}) = \int
\psi_{\rm{f}}^{*}({\mbox{\boldmath $r$}}_{1}
{\mbox{\boldmath $\sigma$}}_{1},{\mbox{\boldmath $r$}}_{2}
{\mbox{\boldmath $\sigma$}}_{2})
J^{\mu}({\mbox{\boldmath $r$}},{\mbox{\boldmath $r$}}_{1}
{\mbox{\boldmath $\sigma$}}_{1},{\mbox{\boldmath $r$}}_{2}
{\mbox{\boldmath $\sigma$}}_{2})\psi_{\rm{i}}
({\mbox{\boldmath $r$}}_{1}{\mbox{\boldmath $\sigma$}}_{1},
{\mbox{\boldmath $r$}}_{2}{\mbox{\boldmath $\sigma$}}_{2})
{\rm{e}}^{{\rm{i}}
{\footnotesize {\mbox{\boldmath $q$}}}
\cdot
{\footnotesize {\mbox{\boldmath $r$}}}
} {\rm d}{\mbox{\boldmath $r$}}
{\rm d}{\mbox{\boldmath $r$}}_{1} {\rm d}{\mbox{\boldmath $r$}}_{2}
{\rm d}{\mbox{\boldmath $\sigma$}}_{1} {\rm d}{\mbox{\boldmath $\sigma$}}_{2} . \label{eq:jq}
\end{equation}
\newline
Eq.~(\ref{eq:jq}) contains three main ingredients: the final-state wave 
function $\psi_{\rm{f}}$, the nuclear current $J^{\mu}$ and the 
two-nucleon overlap integral $\psi_{\rm{i}}$. The bound and
scattering states $\psi_{\rm{i}}$ and $\psi_{\rm{f}}$ are consistently 
derived from an energy--dependent non-hermitian Feshbach-type Hamiltonian
for the considered final state of the residual nucleus. They are 
eigenfunctions of this Hamiltonian at negative and positive energy eigenvalues, 
respectively~\cite{GP,Ox} . 

The same theoretical model for the exclusive (e,e$'$pp) reaction as in 
Ref.\ \cite{GP97} is used, but here an improved treatment of the nuclear 
current and of the two-nucleon overlap integral has been adopted,
as described below. 

In the final-state wave function $\psi_{\rm {f}}$ each of the outgoing 
nucleons interacts with the residual nucleus while the mutual interaction 
between the two outgoing nucleons is neglected. The scattering state is thus 
written as the product of two uncoupled single-particle distorted wave 
functions, eigenfunctions of a complex phenomenological optical potential 
which contains a central, a Coulomb and a spin-orbit term. The effects of an 
isospin-dependent term, to account for charge-exchange final-state 
interactions, were evaluated for the (e,e$'$pp) reaction in Ref.\ \cite{GPex} 
but negligible contributions were obtained in all the situations of practical
interest. Thus this term is neglected here.

The nuclear current $J^{\mu}$ is the sum of a one-body and a two-body
part. The one-body part contains a Coulomb, a convective and a spin term. The
two-body component is derived from the effective Lagrangian of 
Ref.\ \cite{Peccei}, performing a nonrelativistic reduction of the lowest order
Feynman diagrams with one-pion exchange. In this approximation only processes
with $\Delta$-isobar configurations in the intermediate state contribute to 
the (e,e$'$pp) reaction. They produce a completely transverse current, 
${\mbox{\boldmath $J$}}^{\Delta}$. The operator form of 
${\mbox{\boldmath $J$}}^{\Delta}$ was derived in 
Ref.\ \cite{WAGP}. It results from the sum of the contributions due to two 
types of processes, corresponding to the excitation and de-excitation 
part of the current. In the former case, the $\Delta$ is excited by photon 
absorption and then de-excited by pion exchange. The latter process 
describes the time interchange of the two steps, i.e., first excitation of a
virtual $\Delta$ by pion exchange in a NN collision and subsequent
de-excitation by photon absorption. For a pp pair they give~\cite{WAGP}

\begin{equation}
  {\mbox{\boldmath $J$}}^\Delta_{\,I,II}({\mbox{\boldmath $q$}},
  {\mbox{\boldmath $\sigma$}}_1,{\mbox{\boldmath $\sigma$}}_2) =
  \frac{1}{9}\,\gamma\, 
    2{\mbox{\boldmath $\tau$}}^{(2)}_{3} \left( 2{\rm{i}}
    {\mbox{\boldmath $k$}} \mp {\mbox{\boldmath $k$}} \times
   {\mbox{\boldmath $\sigma$}}^{(1)} \right) \times {\mbox{\boldmath $q$}}
   \,\, G_\Delta(\sqrt{s_{I,II}})\,\,
  \frac{{\mbox{\boldmath $\sigma$}}^{(2)}\cdot{\mbox{\boldmath $k$}}}
  {{\mbox{\boldmath $k$}}^2+m^2} \, F(q_{\mu}^{2}),
\label{eq:j1s}
\end{equation}
where ${\mbox{\boldmath $k$}}$ is the momentum of the exchanged pion, $m$ is the pion mass and the 
factor $\gamma$ collects various coupling constants, 
$\gamma=f_{\gamma N\Delta}f_{\pi NN}f_{\pi  N\Delta}/m^3$.
The dipole form factor
\begin{equation}
F(q_{\mu}^2) = \left[ 1-\frac{q_{\mu}^2}{(855\rm{MeV})^2}\right]^{-2}
\end{equation}
takes into account the electromagnetic form factor of the isobar, which
corresponds to the isovector form factor $G_{\rm{M}}^{\rm{V}}$ used in
the static quark model~\cite{Riska}.  The propagator of the resonance, 
$G_\Delta$, depends on the invariant energy $\sqrt s$ of the
$\Delta$, which is different for parts I and II. For the de-excitation current 
$\sqrt{s_{II}}$ is approximated  by the nucleon mass $M$ and 
\begin{equation}
G_\Delta = (M_\Delta-M)^{-1},    \label{eq:ei}
\end{equation}
where $M_{\Delta} = 1232$ MeV.
For the excitation current we use~\cite{WWA}
\begin{equation}
  \sqrt{s_{I}}=\sqrt{s_{NN}}-M,
\label{eq:s1}
\end{equation}
where $\sqrt{s_{NN}}$ is the experimentally measured invariant energy
of the two outgoing nucleons. This gives 
\begin{equation}
  G_\Delta(\sqrt{s_I})=\left({M_\Delta}-\sqrt{s_I}-
\frac{{\rm{i}}}{2}\Gamma_\Delta (\sqrt{s_I}) \right)^{-1},
\label{eq:prop}
\end{equation}
where the decay width of the $\Delta$, $\Gamma_\Delta$, has been taken in the 
calculations according to the parameterization of Ref.\ \cite{BM}.

The two-nucleon overlap integral $\psi_{\rm{i}}$ in (\ref{eq:jq})
contains the information on nuclear structure.
For a discrete final state of the $^{14}$C nucleus, with angular momentum 
quantum numbers $JM$, the relevant part may be expressed in terms of relative
and center-of-mass (CM) wave functions as 
\begin{equation}
\psi_{\rm{i}}({\mbox{\boldmath $r$}}_{1}{\mbox{\boldmath $\sigma$}}_{1},
{\mbox{\boldmath $r$}}_{2}{\mbox{\boldmath $\sigma$}}_{2}) =  \sum_{nlSjNL} \,
c^{i}_{nlSjNL} \, \phi_{nlSj}(r) R_{NL}(R) \, \left[\Im ^{j}_{lS}
(\Omega_r,{\mbox{\boldmath $\sigma$}}_1,{\mbox{\boldmath $\sigma$}}_2) \,
Y_{L}(\Omega_R)\right]^{JM},
\label{eq:ppover}
\end{equation}
where
\begin{equation}
{\mbox{\boldmath $r$}} = \frac{{\mbox{\boldmath $r$}}_{1}-
{\mbox{\boldmath $r$}}_{2}}{\sqrt 2}, \, \, \, \, \, 
{\mbox{\boldmath $R$}} = \frac{{\mbox{\boldmath $r$}}_{1}+
{\mbox{\boldmath $r$}}_{2}}{\sqrt 2}
\end{equation}
are the relative and CM variables. Note that we follow the convention
to denote lower case to relative and upper case to CM coordinate
quantum numbers.
The brackets in (\ref{eq:ppover}) indicate angular momentum coupling of the 
angular and spin wave function $\Im$ of relative motion with the spherical 
harmonic of the CM coordinate to the total angular momentum
quantum numbers $JM$. The CM radial wave function $R$ is that of a harmonic 
oscillator~\cite{BMo},
but the radial wave function $\phi$ of relative motion includes a 
defect funtion in order to account for SRC~\cite{GeAl2}
\begin{equation}
\phi_{nlSj}(r) = R_{nl}(r) + D_{lSj}(r).
\label{eq:def}
\end{equation}
These defect wave
functions were obtained by solving the Bethe-Goldstone equation in momentum
space for $^{16}$O\cite{MS93a}. For the present application these defect
functions were Fourier Bessel transformed into coordinate space.
This is not an exact procedure; the solution of the Bethe-Goldstone
equation yields a non-local correlation operator which cannot strictly
be represented by a local correlation function $D_{lSj}$ of the form
displayed in Eq.~(\ref{eq:def}). 
However, for the $^1S_0$ wave, which is decoupled from other partial waves,
the approximation is quite satisfactory. 
For the higher partial waves of the $pp$ wave function the effect of SRC is
relatively small due to the presence of centrifugal terms.

\begin{figure}[tb]
\epsfysize=6.0cm
\begin{center}
\makebox[16.6cm][c]{\epsfbox{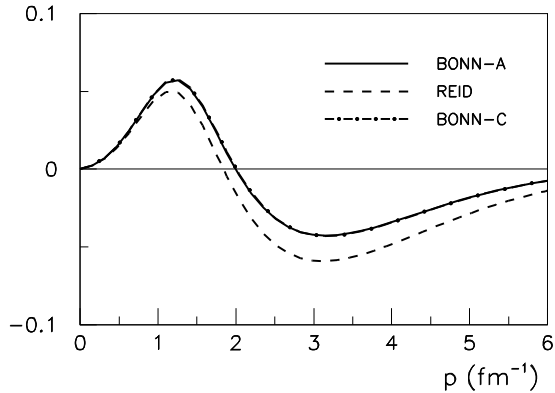}}
\end{center}
\caption[]{Defect functions (see Eq.\ (15)), multiplied by
$p = |{\mbox{\boldmath $p$}}_1-{\mbox{\boldmath $p$}}_2|/2$, calculated
for the ${^1S_0}$ partial wave by solving
the Bethe-Goldstone equation in $^{16}$O, by the method of Ref.\ [28].
Results are plotted for the Bonn-A, Bonn-C and Reid Soft Core potential.
\label{fig:defectAll}
}
\end{figure}

The defect functions for the $^1S_0$ partial wave are displayed in 
Fig.~\ref{fig:defectAll} for the  Bonn-A, Bonn-C and Reid Soft Core 
potentials. One of the objectives of the present study is to investigate
to what extent the differences between these defect functions
are reflected in the calculated cross sections. 

The coefficients $c$ in Eq.~(\ref{eq:ppover}) contain contributions from
a shell-model space which includes the $1s$ up to the $2p1f$ shells. The 
framework within which this is done is basically the same as the one
adopted in a recent calculation of the two-proton removal spectral function
in momentum space~\cite{GeAl2}. The main ingredients of this method are
briefly presented in the next subsection.

\subsection{Structure amplitudes}
The guiding principle followed in the calculation of the structure amplitudes,
which was presented earlier in Ref.\ \cite{GeAl2}, is 
the attempt of treating long-range and short-range
correlations in a separate but consistent way.
The effects of long-range correlations are determined by performing
a nuclear structure calculation  within a shell-model space including
single-particle states which range from the $1s$ up to the $1f2p$ shell.
Limiting the expansion in Eq.~(\ref{eq:ppover}) to this model space
should be sufficient to accomodate long-range correlations.
The effects of the strong short-range components of a realistic
NN interaction, which would scatter the interacting nucleons into much
higher shells, are taken into account by solving the Bethe-Goldstone
equation using a Pauli operator which considers only configurations
outside this model space.
The solution of this Bethe-Goldstone equation yields the residual
interaction of the nucleons inside the model space as well as the
defect functions employed in Eq.~(\ref{eq:def}).
The depletion of filled orbits by SRC
is also incorporated in the shell-model space calculation by the energy
dependence of the G-matrix interaction, which yields an energy dependent
Hartree-Fock term in the self-energy\cite{GeAl1}. The fragmentation of
one-nucleon removal strength is described by two-particle-one-hole and
two-hole-one-particle terms in the self-energy $\Sigma^\ast$ in Tamm-Dancoff
approximation\cite{GeAl1,RGAD96}, with which the Dyson equation for the
one-body propagator 
        \begin{equation}
                g_{\alpha\beta}(\omega)
        =
                g^0_{\alpha\beta}(\omega)
        +
                \sum_{\gamma\delta}
                g^0_{\alpha\gamma}(\omega)
                \Sigma^\ast_{\gamma\delta}(\omega)
                g^{\phantom{0}}_{\delta\beta}(\omega)
        \label{eq:DysonE}
        \end{equation}
is solved. 
In Ref.\ \cite{GeAl1} these dressed propagators were used
to calculate the one-nucleon removal spectroscopic factors for the 
low-energy final states in $^{15}$N. 
The comparison with the results of one-nucleon knock-out experiments is then
a first test of the quality of this ingredient in the calculation of
two-nucleon removal amplitudes. The latter are contained in the Lehmann 
representation of the two-nucleon propagator $G^{II}$
        \begin{eqnarray}
        \lefteqn{%
                G^{II}_{abcd;J}(\omega)
        =}
        \label{eq:g2}
        \\
        &&
                \sum_{n}
                \frac{
                \langle\Psi^A_0|| (a_{\tilde{\beta}} a_{\tilde{\alpha}})_J||
                \Psi^{n,A+2}_J \rangle
                \langle\Psi^{n,A+2}_{J}|| (a^\dagger_{\gamma} 
a^\dagger_{\delta})_J||
                \Psi^A_0 \rangle
                }{ \omega - (E^{n,A+2}_{J} - E^{0,A}_{\phantom{J}}) + i \eta
}
        \nonumber \\
        &&
        -
                \sum_{m}
                \frac{
                \langle\Psi^A_0|| (a^\dagger_{\gamma} a^\dagger_{\delta})_J||%
                \Psi^{m,A-2}_J \rangle
                \langle\Psi^{m,A-2}_{J}||
                        (a_{\tilde{\beta}} a_{\tilde{\alpha}})_J||%
                \Psi^A_0 \rangle
                }{ \omega - (E^{0,A}_{\phantom{J}} - E^{m,A-2}_{J}) - i \eta
}       
        \nonumber \\
        &&
        = 
                \sum_{n}
                \frac{
                Y^{n\ast}_{ab J} Y^{n}_{cd J}
                }{ \omega - (E^{n,A+2}_{J} - E^{0,A}_{\phantom{J}}) + i \eta
}       
        -
                \sum_{m}
                \frac{
                X^{m\ast}_{cd J} X^{m}_{ab J}
                }{ \omega - (E^{0,A}_{\phantom{J}} - E^{m,A-2}_{J} ) - i \eta }      
        \nonumber
        \;.
        \end{eqnarray}
The symbols $\langle\ldots||\ldots||\ldots\rangle$ represent the
reduced matrix elements%
\cite{Ed57,BG77,FW71}
of the two-nucleon removal and addition tensor operators that
are constructed by the angular momentum coupling of two one-nucleon addition
and removal tensors $a^\dagger_{\alpha}$ and $a_{\tilde{\alpha}}$, where
$a_{\tilde{\alpha}}=(-)^{j_\alpha-m_\alpha}a_{-\alpha}$ is the
time reverse of $\alpha$; 
$-\alpha$ denotes $\{n_\alpha, l_\alpha, j_\alpha, -m_\alpha\}$ and 
$a$ denotes basis states without the magnetic quantum number: 
$a=\{n_\alpha, l_\alpha, j_\alpha\}$.

The two-nucleon propagator is obtained by solving,
within the shell-model space,
the Bethe-Salpeter equation\cite{FW71,AGD63} for the two-nucleon
propagator $G^{II}$
        \begin{eqnarray}
        \lefteqn{%
                G^{II}_{\alpha\beta\gamma\delta}
                         (t_1, t_2, t_3, t_4)
        =} && \label{eq:BSEpp} \\
        &&
        i\left[
                g_{\alpha\gamma}(t_1-t_3)
                g_{\beta\delta}(t_2-t_4)
        -
                g_{\alpha\delta}(t_1-t_4)
                g_{\beta\gamma}(t_2-t_3)
        \right]
        \nonumber \\
        &&
        - 
                \int\limits_{-\infty}^{\infty}
                {\rm d} t'_1 {\rm d} t'_2 {\rm d} t'_3 {\rm d} t'_4
               \sum_{\mu\nu\kappa\lambda}
	       \nonumber \\
	       &&
	       \left[
		     g_{\alpha\mu}(t_1-t'_1)
		     g_{\beta\nu}(t_2-t'_2)
		\right]
			\Gamma^{pp}_{\mu\nu\kappa\lambda}
               (t'_1, t'_2, t'_3, t'_4)
                G^{II}_{\kappa\lambda\gamma\delta}
                (t'_3, t'_4, t_3, t_4)
        \nonumber
        \;,
        \end{eqnarray}
where $\Gamma$ denotes the irreducible effective particle-particle interaction,
which is here approximated by the G-matrix interaction which contains only
propagation of particles outside the chosen model space.

In the calculation of Ref.\cite{GeAl1} the spectroscopic factor for the 
removal of {\em one} nucleon from the $p$ shell of $^{16}$O  turned out to be
reduced by a factor 0.75 as compared with the independent-particle shell model.
This is still about 10\% 
larger than the factor 0.65 deduced from experiments\cite{Leu94}. We 
decided not to replace the calculated spectroscopic factor by the 
experimental ones in the dressed propagators. This means that the
{\em two}-nucleon removal amplitudes that we obtain in the RPA with these 
dressed propagators\cite{GeAl2}
may be too large as well.
This observation applies mostly to the noninteracting part of the
two-particle spectral function
represented by the first contribution to the two-nucleon
propagator in Eq.\ (\ref{eq:BSEpp}).
This term also yields a spurious contribution to the
cross section for the one-body current contributions at small
momenta~\cite{GeAl2}.
The issue of interest here involves the effect of SRC which appear at higher
momenta and the problem of spuriosity is not important.
The overestimate may be much less severe for the interacting part of the
spectral function (second term in Eq.\ (\ref{eq:BSEpp})) which
yields the genuine SRC contribution to the cross section.
In addition, such a factor, representing this overestimate,
will be roughly the same for all the low-energy
amplitudes involving removal of two protons from the $p$ shell and therefore
this uncertainty cancels in the comparison of {\em relative}
magnitudes of amplitudes and cross sections for the low-energy final
states in $^{14}$C.

The shell-model two-proton removal amplitudes are expanded in terms of 
relative and CM wave functions for the initial state of the 
knocked-out pair. Summation over the contribution of the various 
configurations yields the coefficients $c$ in Eq.\ (\ref{eq:ppover}):
        \begin{eqnarray}
\lefteqn{               c^{i}_{nlSjNL}
        =} \nonumber \\   
        &&\sum_{ab} \sum_\lambda
                (-)^{L+\lambda+j+S}
                (2\lambda + 1) \hat{j} \hat{S} \hat{j}_a \hat{j}_b
                \left\{\matrix{l_a & l_b & \lambda \cr
                               s_a & s_b & S \cr
                               j_a & j_b & J\cr}
                \right\}  
                \langle n l N L \lambda|
                n_a l_a n_b l_b \lambda\rangle
                \left\{\matrix{L & l & \lambda \cr
                               S & J & j \cr}
                \right\} X^{i}_{ab J} 
        \;,
        \label{eq:Cab}
        \end{eqnarray}
with the notation
$\hat{j}=\sqrt{2j+1}$ 
and the nine-j and six-j symbols come from the angular momentum recouplings
involved\cite{GeAl2,Ed57}.

The most important amplitudes are listed in 
Table~\ref{tab:Amplitudes}. It is instructive
to note that for these low-lying positive parity states the relative
$^1S_0$ wave is combined with a CM  $L=0$ (for $0^+$) or
$L=2$ (for $2^+$) wave, while the relative $^3P$ waves occur always 
combined with a $L=1$ CM wave function. This was the basis
of the global analysis of the experimental cross section in terms of
$^1S_0$ and $^3P$ removal contributions in Ref.\ \cite{OnAl}.
The amplitudes for the $0^+$ states are presented at some length to
illustrate the importance of the pairing interaction which mixes the
shell-model configurations. Without this interaction, the lowest state
would just correspond to the removal of two (dressed) protons from the
$p_{\frac{1}{2}}$ shell and the excited $0^+$ state to the removal from
the $p_{\frac{3}{2}}$ shell. In that case the $^1S_0$ removal cross
section would be twice as large for the excited state as for the ground
state. Due to the residual interaction the ground state becomes the strongest
for $^1S_0$ removal, not only due to the coherent superposition
of the $p$ shell configurations but also the deep $1s$ shell and the
higher $sd$ and $pf$ major shells contribute. The contribution from
these higher shells is much smaller for the $2^+$ states and completely
negligible for the $1^+$ state.

Another point to be mentioned is that the squares of the amplitudes
add up to only about 0.6, as to be expected on the basis of the
products of two one-nucleon removal spectroscopic factors
$(0.75)^2$.

\begin{figure}[tb]
\epsfysize=14.0cm
\begin{center}
\makebox[16.6cm][c]{\epsfbox{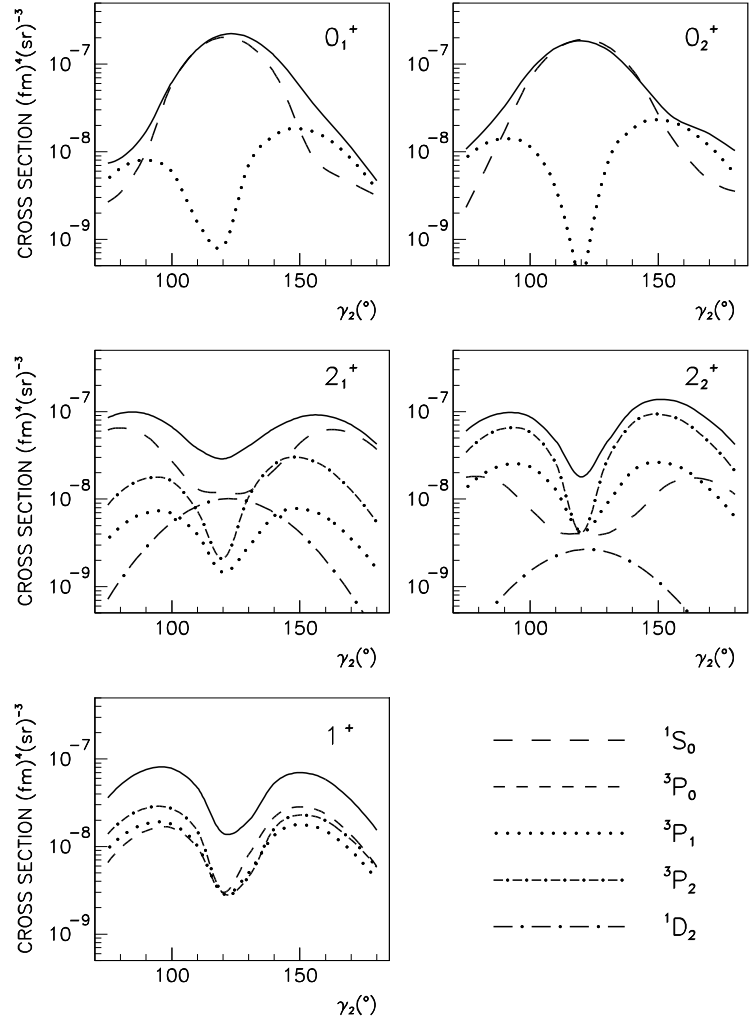}}
\end{center}
\caption[]{The differential cross section  of the $^{16}$O(e,e$'$pp)
reaction as a function of the angle
$\gamma_{2}$ for the transitions to the low-lying
states in $^{14}$C: $0^+_1$ ($E_{2\rm{m}} = 22.33$ MeV), $0^+_2$
($E_{2\rm{m}} = 32.08$ MeV), $2^+_1$ ($E_{2\rm{m}} = 30$ MeV),
$2^+_2$ ($E_{2\rm{m}} = 35.47$ MeV), $1^+$ ($E_{2\rm{m}} = 33.64$ MeV).
$E_{0} = 584$ MeV, $\omega = 212$ MeV, $q = 300$ MeV/$c$, $T'_{1} = 137$ MeV
and $\gamma_{1} = -30^{\rm{o}}$. The defect functions for the Bonn-A
potential and the optical potential of Ref.\ [37] are used. Separate
contributions of different relative partial waves are drawn. The contribution
of the  $^1D_2$ partial wave is very small for the $0^+$ states and omitted
from the figure. The solid lines give the total cross sections resulting from
the contributions of all the relative states.
\label{fig:rel584}
}
\end{figure}


\section{Two-proton knock-out cross sections}

\subsection{Relative magnitude of the contributions from one-body and
two-body currents}
\begin{figure}[tb]
\epsfysize=6.0cm
\begin{center}
\makebox[16.6cm][c]{\epsfbox{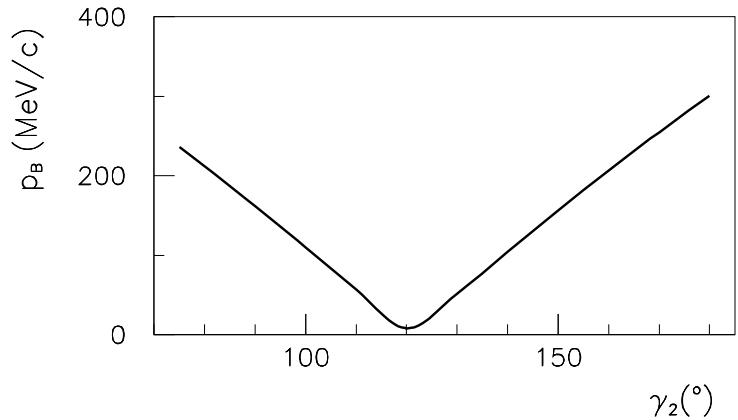}}
\end{center}
\caption[]{The recoil momentum $p_{\rm{B}}$ as a function of $\gamma_{2}$
in the same kinematics as in Fig.~\ref{fig:rel584}.
\label{fig:pb}
}
\end{figure}

Of major interest in the (e,e$'$pp) studies is the question whether one may 
clearly identify the contributions from one-body and two-body currents and 
thereby study them separately. The part involving the one-body current is 
expected to provide then an opportunity to probe SRC. These SRC, induced by 
the repulsive NN interaction, with a range of typically 0.5~fm, will 
strongly affect the relative $^1S_0$ wave function, but the short-range 
repulsion will have only a minor impact on the higher partial waves. For this 
reason a first inspection of the experimental data from NIKHEF has been made 
in Ref.\ \cite{OnAl} to estimate the relative contribution of $^1S_0$ and 
$^3P$ pair knock-out in the cross sections for the lowest states of $^{14}$C. 
This estimate was based on the comparison
of a simple factorization approximation of the cross section with the observed
distribution of CM momenta of the knocked-out pairs (see also Ref.\
\cite{Ry2}). Here we
present the separate contributions of the $^1S_0$, $^3P_j$ and
$^1D_2$ relative partial waves to the $^{16}$O(e,e$'$pp) cross sections for
the low-lying states in $^{14}$C. They are displayed in Fig.~\ref{fig:rel584} 
for a specific kinematical setting that is included in the aforementioned 
NIKHEF data, with $E_0 =584$ MeV, $\theta = 26.5^{\rm{o}}$, $\omega = 212$ 
MeV, and $q= 300$ MeV/$c$. The kinetic energy of the first outgoing proton 
$T'_{1}$ is 137 MeV. The missing energy
$E_{2\rm{m}} =\omega-T'_{1}-T'_{2}-T'_{\rm{B}}$, where $T'_{2}$ and
$T'_{\rm{B}}$ are the kinetic energies of the second outgoing proton and
of the residual nucleus, respectively, has been taken in the calculations,
for each transition, from a comparison with the experimental spectrum
of $^{14}$C\ \cite{Ajz1315} but for the $2^+_2$ state, unidentified in the
 experimental spectrum, from the calculation of Ref.\ \cite{GeAl2}. The
angle $\gamma_{1}$ is $-30^{\rm{o}}$, on the opposite side of the outgoing
electron with respect to ${\mbox{\boldmath $q$}}$.
Changing the angle $\gamma_{2}$ on the other
side the cross section  can be explored at different values of the recoil
momentum
$ {\mbox{\boldmath $p$}}_{\rm{B}}$.
The relationship between $\gamma_{2}$ and $p_{\rm{B}}$
is shown in Fig.~\ref{fig:pb}  for the transition to the ground state of 
$^{14}$C. Only small differences are obtained for the other states, owing to 
the different value of the missing energy.

In a factorized approach, where final-state interaction is neglected,
${\mbox{\boldmath $p$}}_{\rm{B}}$ is opposite to the total momentum
of the initial nucleon
pair. Thus in this approach the shape of the recoil momentum distribution is
determined by the CM orbital angular momentum $L$ of the knocked-out
pair. This feature is not spoiled by final state interaction, which modifies
the pair momentum. In fact in Fig.~\ref{fig:rel584} the shapes of the angular 
distributions
for different transitions and separate contributions of different relative
states are determined by the corresponding values of $L$, indicated in 
Table~\ref{tab:Amplitudes}. The shape of the total result is driven by the 
component which gives the major contribution. Due to final-state interaction 
there is interference of different partial waves in the total cross section. 
In some cases it can be important, but in certain kinematical regions 
this is of minor importance, because there either one is much stronger than the
other.

The figures show that the cross section for the $0^+$ ground state, for the
$0^+_2$ and to a lesser extent also for the $2^+_1$ state of $^{14}$C,
receive a
major contribution from the $^1S_0$ knock-out, as opposed to the higher lying
states $1^+$, where only $^3P$ waves contribute, and $2^+_2$, where the
$^3P$ waves are more prominent. This feature is in agreement with the
experimental findings of Ref.\ \cite{OnAl}.
The defect functions used in the calculations of Fig.~\ref{fig:rel584} 
were those of the Bonn-A potential\cite{GeAl2}. The results for the Reid Soft 
Core potential have a similar qualitatitive behaviour for this case and 
therefore are not presented here. Calculations with the Bonn-C potential have 
not been performed, but from the shape of the defect functions shown in 
Fig.~\ref{fig:defectAll} we do not expect any significant difference with 
respect to the results obtained with the Bonn-A potential. 

\begin{figure}[tb]
\epsfysize=10.0cm
\begin{center}
\makebox[16.6cm][c]{\epsfbox{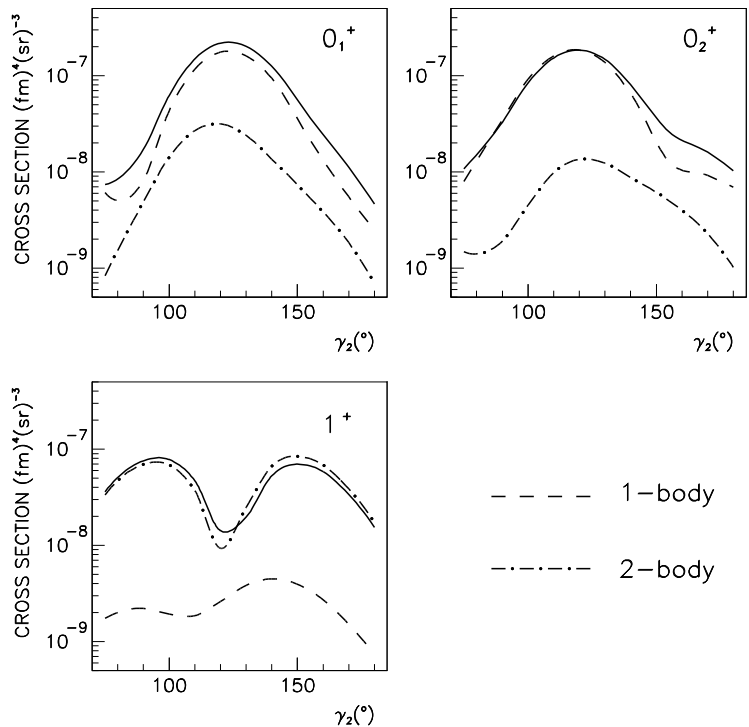}}
\end{center}
\caption[]{The differential cross section  of the $^{16}$O(e,e$'$pp)
reaction as a function of $\gamma_{2}$ for the transitions to the $0^+_1$,
$0^+_2$ and $1^+$ states in $^{14}$C in the same kinematics as in
Fig.~\ref{fig:rel584}. Defect functions and optical potential as in
Fig.~\ref{fig:rel584}. Separate contributions of the one-body and of the
two-body $\Delta$-current are shown. The solid lines are the same as in
Fig.~\ref{fig:rel584}. \label{fig:curr584}
}
\end{figure}

\begin{figure}[tb]
\epsfysize=10.0cm
\begin{center}
\makebox[16.6cm][c]{\epsfbox{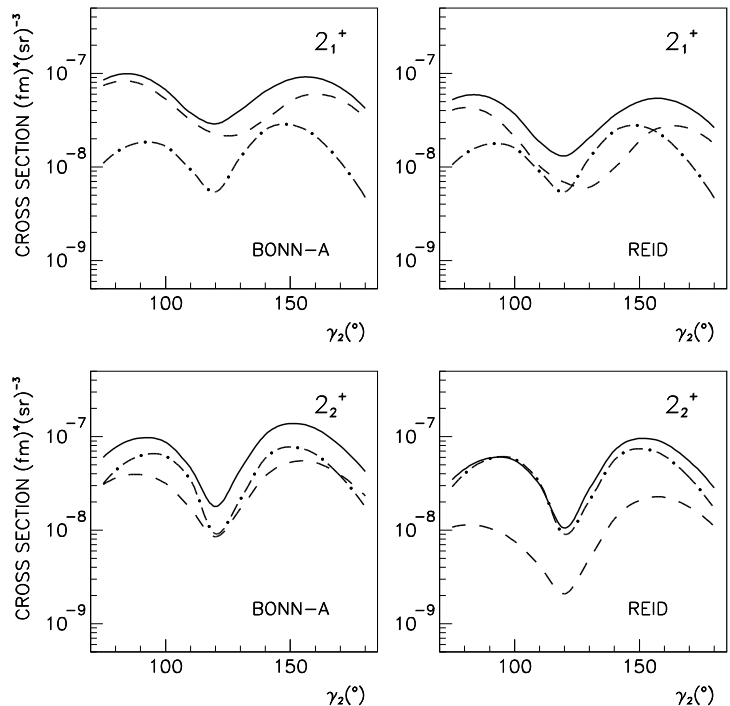}}
\end{center}
\caption[]{The differential cross section  of the $^{16}$O(e,e$'$pp)
reaction as a function of $\gamma_{2}$ for the transitions to the $2^+_1$ and
$2^+_2$ states in $^{14}$C in the same kinematics as in
Fig.~\ref{fig:rel584}. Separate contributions of the one-body and of the
two-body $\Delta$-current are shown for the defect functions calculated with
the Bonn-A and Reid potentials. The solid lines give the total cross
sections resulting from the sum of the one-body and of the two-body
$\Delta$-current.
Line convention and optical potential as in Fig.~\ref{fig:curr584}.
\label{fig:curr2584}
}
\end{figure}

As already mentioned, one may hope that the one-body current and thus
correlations yield the dominant contribution to the cross section in some
kinematical regions when the knocked-out pair is in a $^1S_0$ state. The
knock-out of $^3P$ and higher partial waves will proceed mainly through
the two-body $\Delta$-current.
To illustrate to what extent our calculations support these expectations,
we have plotted in Figs.~\ref{fig:curr584} and \ref{fig:curr2584} the separate 
contributions from the one-body and two-body current to the same total cross 
section as in Fig.~\ref{fig:rel584}.
For the $0^+$ states the contribution of the one-body current is
much larger than that of the two-body current and the angular dependence
has the $s$-wave shape typical of the $^1S_0$ contribution for these states.
The results with the Reid defect functions have a similar shape but are a 
factor two smaller . In fact the range of 
relative momenta $p_{rel} = |{\mbox{\boldmath $p$}}_1-
{\mbox{\boldmath $p$}}_2|/2$ probed in this region is 
$\approx$~1.5 fm$^{-1}$, where the ratio of the Bonn-A and Reid $^1S_0$ 
defect functions is $\approx$~1.4, which gives a factor two in the cross 
section.  For larger values of the recoil momentum the $s$ wave 
becomes smaller, while the $p$ wave becomes relatively more important.
In the range of angles between 
100$^{\circ}$ and 140$^{\circ}$, where the
recoil  momentum is small, one may therefore probe correlations in
the relative $^1S_0$ wave function.                                    

In sharp contrast to the $0^+$ states is the situation for the $1^+$ state.
It is only reached by the knock-out of $^3P$ pairs and,  as expected, the 
two-body current gives here by far the dominant contribution to the cross
section. It will therefore be interesting to identify this cross section 
for the $1^+$, which is known to be at 10.3 MeV excitation energy.

For the $2^+_1$ state we find that the one-body current gives a larger
contribution than the two-body current, as opposed to the situation for
the $2^+_2$ state. This may be traced back to the large contribution of
the $^1S_0$ partial wave for the $2^+_1$, as was shown in 
Fig.~\ref{fig:rel584}. 
For the $2^+_2$ especially $^3P_2$ dominates. However, the predicted
dominance of the one-body contribution to the $2^+_1$ cross section depends
on the defect functions used. This is shown by the comparison between the
results obtained with defect functions from the Bonn-A and from the Reid 
potential in Fig.~\ref{fig:curr2584}. With the Reid defect functions the 
one-body current contribution is almost a factor two smaller than for the 
Bonn-A defect functions. This is not a general statement, but it turns out to 
be the case for the present kinematics. The cross section calculated with 
the two-body current is, as expected, only slightly affected by the choice of 
the defect functions. With the Reid defect functions the amplitudes from one- 
and two-body currents become of about the same size for the $2^+_1$ state and 
the shape of the total cross section is determined by the interference of 
the two contributions. A similar result is obtained with the Bonn-A defect 
functions for the $2^+_2$ state. 

\begin{figure}[tb]
\epsfysize=6.0cm
\begin{center}
\makebox[16.6cm][c]{\epsfbox{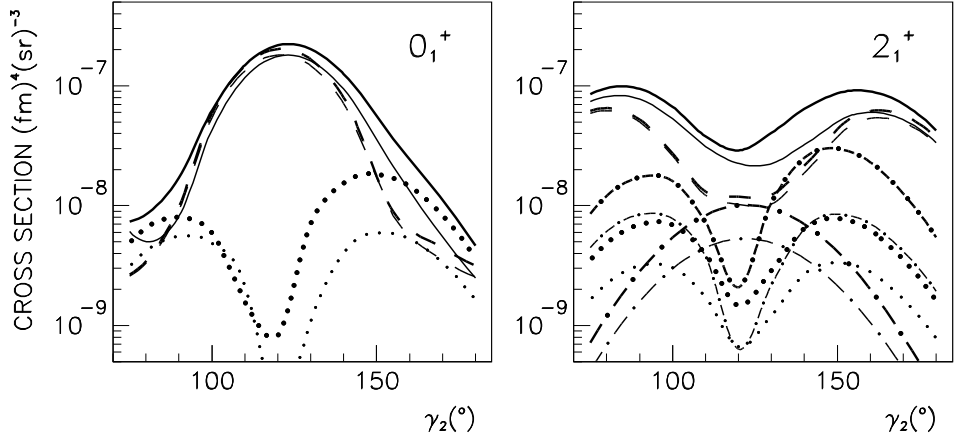}}
\end{center}
\caption[]{The differential cross section  of the $^{16}$O(e,e$'$pp)
reaction as a function of $\gamma_{2}$ for the transitions to the $0^+_1$ and
$2^+_1$ states in $^{14}$C in the same kinematics and with same line
convention as in Fig.~\ref{fig:rel584}. Defect functions and optical potential
as in Fig.~\ref{fig:rel584}. The thin, mostly lower lines are calculated with
the one-body current only.
\label{fig:delta584}
}
\end{figure}

Next, we show explicitly how the amplitudes for knock-out from $^1S_0$ and 
higher partial waves are influenced by the $\Delta$-current. This is plotted 
in Fig.~\ref{fig:delta584} for the $0^+_1$ and $2^+_1$ states. The figures 
illustrate that indeed the $^1S_0$ knock-out amplitude is relatively
little affected by the inclusion of the $\Delta$-current, while this
two-body current is a major factor in the knock-out of $^3P$ and $^1D$
waves. This is a general result that has been obtained also in other 
kinematical situations. It can be understood if we consider the different role 
of the excitation and de-excitation part of the $\Delta$-current.
The excitation
current, which has the energy-dependent $\Delta$-propagator of
Eq.~(\ref{eq:prop}), gives for energy transfer above 150 MeV the dominant
contribution of the $\Delta$-current on $^3P$ and $^1D$ waves. On a $^1S_0$ pp
pair the contribution of the excitation current is strongly reduced~\cite{WNA}
and becomes in our calculation of about the same size or even smaller than that
of the de-excitation current, which is generally small. 
Thus the contribution of
the $\Delta$-current is generally small on a $^1S_0$ pp pair, while it is
dominant on $^3P$ and $^1D$ pp pairs. The contribution of the $^1D$ waves to
the total cross section is generally very small. So the proper place to study
the two-body $\Delta$-current in the (e,e$'$pp) reaction is where the $^3P$
knock-out dominates, as in the $1^+$ and $2^+_2$ states, while  SRC should be
studied in the lowest states, where $^1S_0$ knock-out dominates. Whether indeed
one of these is dominant can be verified by inspection of the pair momentum
distribution, as was illustrated in Ref.\ \cite{OnAl}.

\begin{figure}[tb]
\epsfysize=12.0cm
\begin{center}
\makebox[16.6cm][c]{\epsfbox{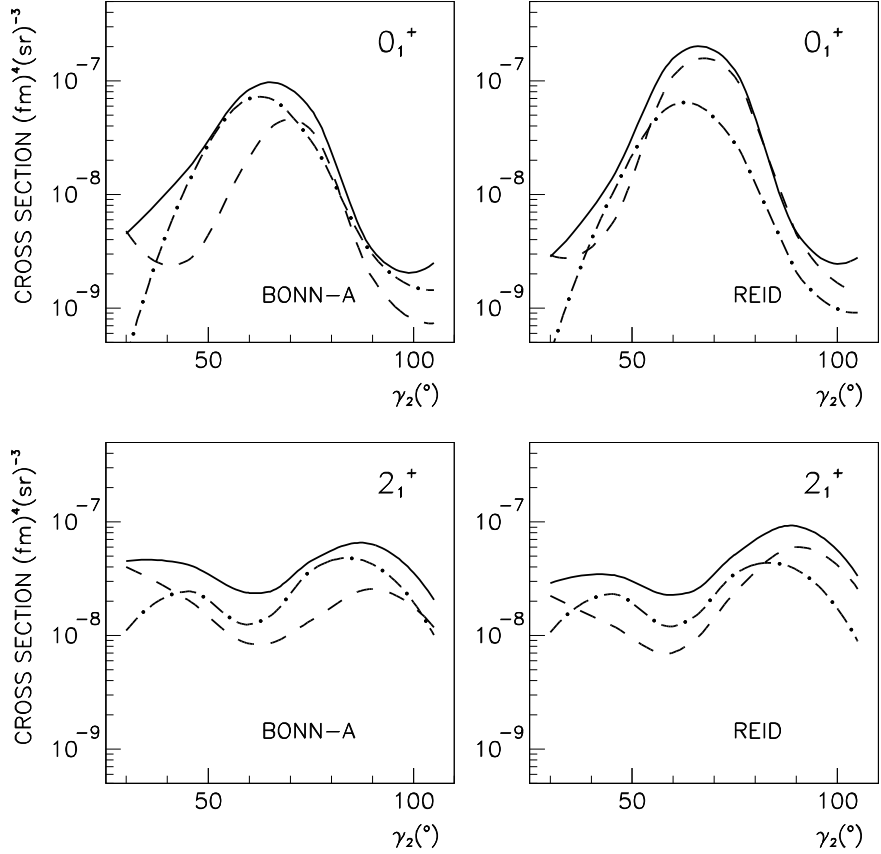}}
\end{center}
\caption[]{The differential cross section  of the $^{16}$O(e,e$'$pp)
reaction as a function of $\gamma_{2}$ for the transitions to the $0^+_1$ and
$2^+_1$ states in $^{14}$C, now in the same kinematics as in
Ref.\ [12]: $E_{0} = 475$ MeV, $\omega = 212$ MeV, $q = 268$ MeV/$c$,
$T'_{1} = 68$ MeV and $\gamma_{1} = 79.2^{\rm{o}}$. Line convention as in
Fig.~\ref{fig:curr584}.
\label{fig:curr475}
}
\end{figure}
                               
\subsection{Dependence on the NN potential and on the probed range of 
momenta}
In the discussion of Fig.~\ref{fig:curr2584} it was already indicated that 
especially the cross sections due to correlations and the one-body current are 
sensitive to the defect functions and thereby to the NN potential from which 
these were derived. For the range of relative momenta probed in the cross 
sections of Figs.~\ref{fig:rel584}-\ref{fig:delta584}, the $^1S_0$ defect 
function of the Bonn-A potential is larger than that of Reid. In different 
kinematical situations it may be just the opposite. This appears to be the 
case for instance in the kinematics of Ref.~\cite{GP97}. In 
Fig.~\ref{fig:curr475} it is shown that with that kinematics the
contribution of the one-body current to the cross section for $0^+_1$ and
$2^+_1$ is for most angles larger for Reid than for Bonn-A. The range of
relative momenta probed here is on the average higher than in 
Figs.~\ref{fig:rel584}-\ref{fig:delta584}. 
Around $\gamma \approx$ 65$^{\circ}$ the $0^+_1$ cross section is probed with
$p_{rel} \approx$ 2.1 fm$^{-1}$. For the $2^+_1$ state the maximum around
$\gamma \approx$ 90$^{\circ}$ corresponds to $p_{rel} \approx$ 2.2 fm$^{-1}$.
                                                                             
For really high relative momenta, above $p_{rel} \approx$ 3 fm$^{-1}$, the
contribution of the one-body current to the cross section will become
systematically about a factor two larger for Reid than for the Bonn
potentials. This is clear from the momentum dependence of the $^1S_0$
defect wave functions that were shown in Fig.~\ref{fig:defectAll}. These might 
be probed in
future experiments at TJNAF. Another possibility to discriminate between
these potentials could be provided by the separation of structure functions.
We discuss an example of this in the next subsection.

\begin{figure}[tb]
\epsfysize=12.0cm
\begin{center}
\makebox[16.6cm][c]{\epsfbox{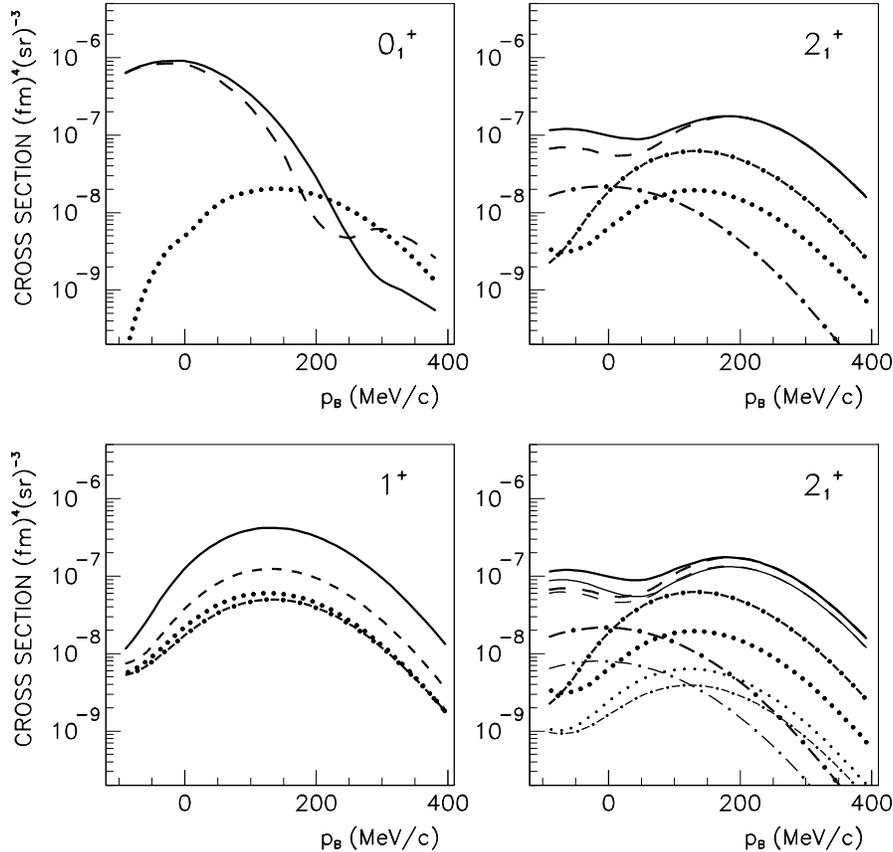}}
\end{center}
\caption[]{The differential cross section  of the $^{16}$O(e,e$'$pp)
reaction as a function of the recoil momentum
$p_{\rm{B}}$ for the transitions to the $0^+_1$,
$2^+_1$  and $1^+$ states in $^{14}$C in a super-parallel
kinematics ($\gamma_{1} = 0^{\rm{o}}$, $\gamma_{2} = 180^{\rm{o}}$)
with $E_{0} = 855$ MeV, $\omega = 215$ MeV and $q = 315.89$ MeV/$c$. The
recoil-momentum distribution is obtained changing the kinetic energies of the
outgoing protons. Line convention, optical potential and defect functions as in
Fig.~\ref{fig:delta584}. Positive (negative) values of $p_{\rm{B}}$ refer
to situations where ${\mbox{\boldmath $p$}}_{\rm{B}}$ is
parallel (antiparallel) to ${\mbox{\boldmath $q$}}$.
\label{fig:mainz}
}
\end{figure}

\subsection{Separation of the structure functions $f_{00}$ and $f_{11}$ in
super-parallel kinematics}
The experimental separation of structure functions appears in general extremely
complicated. The so-called super-parallel kinematics, where the knocked-out
protons are detected parallel and anti-parallel to the transferred momentum
${{\mbox{\boldmath $q$}}}$, is favored by the fact that only two structure
functions, $f_{00}$
and $f_{11}$,  contribute to the cross section, as in the inclusive electron
scattering, and, as in that case, they can in principle be separated by a
Rosenbluth plot~\cite{GP}. This kinematical setting has been realized in a
recent experiment at Mainz\cite{Rosner}. In this experiment, with an energy
resolution of less than 1 MeV, different final states can be separated in the
excitation-energy spectrum of the residual nucleus, in particular the $2^+$
states at 7.01 and 8.32 MeV. To compare the experimental results with our
calculations, however, these two states should be considered as one state, the
$2^+_1$, which is split up by the coupling to excitations of the $^{16}$O
core, that are very complicated and not included in our description.

In Fig.~\ref{fig:mainz} we display the cross sections for the $0^+_1$ ground 
state and the $2^+_1$ and $1^+$ states in the super-parallel kinematics of the 
Mainz experiment, where 
$\gamma_{1} =0^{\rm{o}}$, $\gamma_{2} =180^{\rm{o}}$,
$E_0 =855$ MeV, $\theta = 18^{\rm{o}}$, $\omega = 215$ MeV and
$q= 315.89$ MeV/$c$. The kinetic energy of the outgoing protons is changed in 
the calculations in order to explore different values of $p_{\rm{B}}$.
The figures show the decomposition into the different partial waves of the
knocked-out pair. The recoil momentum distributions are similar to those shown
in Fig.~\ref{fig:rel584}. The shapes of the different relative waves are 
determined by the
corresponding value of $L$. The $0^+_1$ state is dominated for low values of
$p_{\rm{B}}$, up to about 150 MeV/$c$, by $^1S_0$ knock-out. At higher recoil
momenta the contributions of $^1S_0$ and $^3P_1$ knock-out become of the same
order. We observe in this region that the total cross section may be lower
than that given by the two separate contributions of $^1S_0$ and $^3P_1$
states, owing to the negative interference of the two contributions.
The $2^+_1$ state is dominated over the whole range of recoil momenta by
$^1S_0$ knock-out, whose contribution is about a factor four larger than that of
the other relative states. So in this kinematics the $2^+_1$ seems to offer
the best opportunity to study correlation effects.

We do not display a decomposition into contributions from the one-body and
two-body currents here, because the results are conceptually similar to those
given in Figs.~4 and 5 and indicate the dominance of the one-body current for
the $0^+$ and the $2^+_1$ states and of the $\Delta$-current for the $1^+$
state. Moreover, the figures look quite similar to the ones shown here,
{\em i.e.} the contribution of the one-body current is practically the same as
that of the $^1S_0$ removal while higher partial waves come almost exclusively
from the two-body current. This is illustrated explicitly for the $2^+_1$
state in the last frame of Fig.~\ref{fig:mainz}.

Essentially the same results as those shown in Fig.~\ref{fig:mainz}, for the 
Bonn-A defect functions, are obtained with those of the Reid potential.  In 
the latter case the one-body part is about 20\% smaller, but otherwise the 
distributions are quite similar to those of Fig.~\ref{fig:mainz}.
                                
\begin{figure}[tb]
\epsfysize=12.0cm
\begin{center}
\makebox[16.6cm][c]{\epsfbox{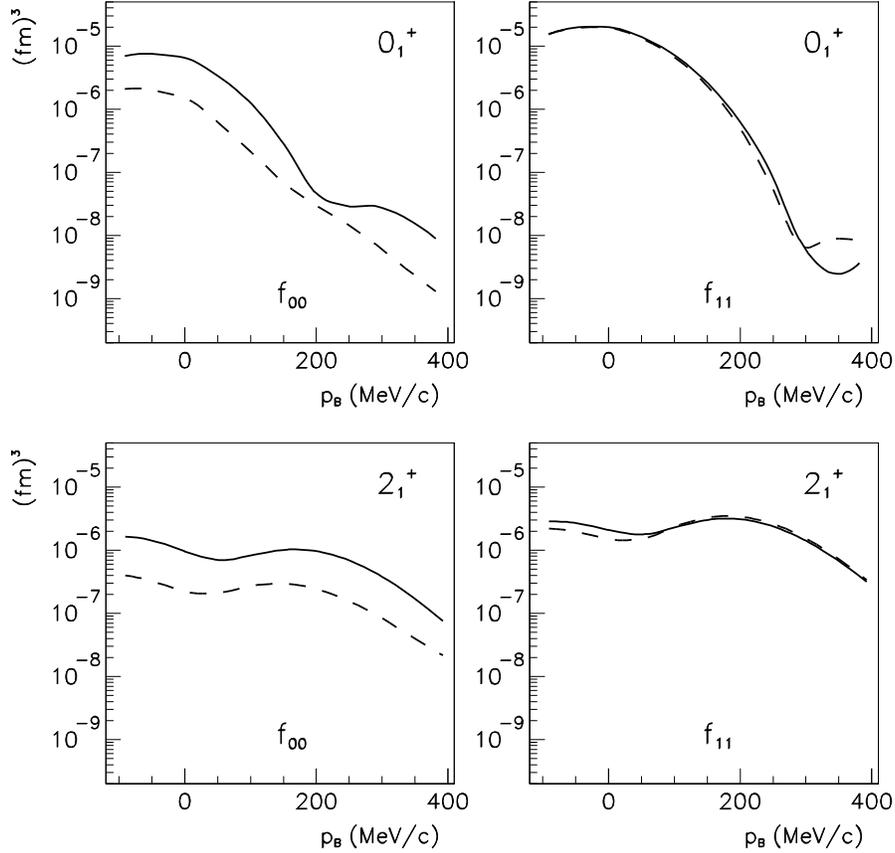}}
\end{center}
\caption[]{The structure functions $f_{00}$ and $f_{11}$ of the
$^{16}$O(e,e$'$pp) reaction as a function of $p_{\rm{B}}$ for the
transitions to the $0^+_1$ and $2^+_1$ states in $^{14}$C in the
super-parallel kinematics of Fig.~\ref{fig:mainz}. Optical potential as in
Fig.~\ref{fig:mainz}. The solid and dashed lines are calculated with the
defect functions of the Bonn-A and Reid potentials, respectively.
\label{fig:mainzff}
}
\end{figure}
 
A large difference between the results with the defect functions of Bonn-A and 
Reid potentials appears when a splitting into contributing structure functions 
$f_{00}$ and $f_{11}$ is
made. These results are plotted in Fig.~\ref{fig:mainzff}. The transverse
structure function $f_{11}$ appears to be insensitive to the defect functions.
This is only partly due to the effect of the $\Delta$-current. On the
contrary the longitudinal structure function $f_{00}$, which is entirely due to
the one-body current and thus to correlations, is much more sensitive than
$f_{11}$ and the cross section to the choice of the defect functions. In this
kinematics $f_{00}$ calculated with the Bonn-A defect functions is typically
four times larger than calculated with the Reid defect functions. However, the
experimental separation of the structure functions may be difficult, since
$f_{11}$ is almost an order of magnitude larger than $f_{00}$ for the
$0^+_1$ ground state. Also for the $1^+$, not shown in the figure, the
$f_{11}$ structure function is found to be roughly five times $f_{00}$
with the Bonn-A defect functions and about twenty times $f_{00}$ with the
Reid defect functions. Somewhat more favorable is the situation for the
$2^+_1$ state, since here $f_{11}$ is only three times larger than $f_{00}$
at $p_B \approx $ 150 MeV/c, if the prediction with the Bonn-A defect
functions turns out to be correct. So this state may offer the best
opportunity to determine the longitudinal structure function $f_{00}$
experimentally.
                             
\section{Summary and conclusions}
This work represents a combination of state of the art reaction description
of the (e,e$'$pp) reaction together with a corresponding calculation of the
two-nucleon spectral function to produce results for cross sections measured
at NIKHEF and Mainz for the $^{16}$O target.
The description of the reaction includes both one- and two-body contributions
to the electromagnetic current. The treatment of final state interactions
of the detected protons incorporates distortions (through an optical
potential) for the individual particles but not their mutual interaction.
Although the latter is expected to be unimportant for the cases of interest,
this issue should be further studied in the future.
The description of the two-body current involves a proper treatment of the
dynamics of the intermediate excitation of the $\Delta$-isobar before or
after the absorption of the virtual photon.
The two-nucleon spectral function (or two-nucleon overlap function)
has been obtained from a two-step procedure.
The calculation of long-range correlations is performed in a shell-model
space large enough to incorporate the corresponding collective features
which influence the pair removal amplitude.
The single-particle propagators used for this dressed Random Phase
Approximation description of the two-particle propagator also include the
effect of both long- and short-range correlations.
In the second step that part of the pair removal amplitudes which describes
the relative motion of the pair, is supplemented by defect functions
obtained from the same G-matrix which is also used as the effective interaction
in the RPA calculation.

An important conclusion in this work concerns the predicted selectivity
of the (e,e$'$pp) reaction involving discrete final states.
Whereas the lowest $0^+$ and $2^+$ in $^{14}$C are predominantly reached by
the removal of a $^1S_0$ pair other states at higher excitation energy
mostly involve $^3P$ removal.
The latter pair removal proceeds primarily via intermediate excitation of
the $\Delta$-isobar whereas the former is dominated by the one-body current
mechanism.
This feature is responsible for the calculated sensitivity in the cross
sections to the treatment of short-range correlations where $^1S_0$ removal
dominates.
Short-range correlations induced by the Bonn- or Reid-potential  may  each yield
larger cross  sections than the other in certain kinematical domains.
As a result,  one will be able to study short-range correlations in this
reaction successfully provided a sufficiently large set of kinematical
conditions is explored including those available at TJNAF.
The most promising extraction of the effect of short-range correlations
shows  up  in the longitudinal structure function which may be studied in
the so-called super-parallel kinematics.
Our study demonstrates that an intelligent choice of kinematics
in exclusive (e,e$'$pp) experiments should allow the separation of
the effects due to isobar currents and SRC for two nucleons with
isospin $T = 1$.
This success gives rise to the hope that a similar separation between
two-body currents and SRC might also be possible in (e,e$'$pn) reactions.
In this case one has to consider the competition between meson-exchange
currents and SRC.
The emission of a $pn$ pair, however, probes the SRC for $T = 0$ which
are even stronger due to the presence of the tensor force.

\acknowledgments
This work forms part of the research program of the Foundation of
Fundamental Research of Matter (FOM), which is financially supported by
the Netherlands' Organization for Scientific Research (NWO). Additional
support is provided by the U.S. National Science Foundation under Grant No.
PHY-9602127 and the DFG Programm ``Untersuchung der hadronischen Struktur
von Nukleonen und Kernen mit elektromagnetischen Sonden'' under
Grant No.  Wa 728/3-1 (Germany).
\vfil\eject

\vfil\eject
\begin{table}
\begin{tabular}{|cccccc|}
\hline
& $n$ & $N$ & $\rho$  & $0^+_1$  & $0^+_2$\\
$^1S_0 ; L = 0$ & 0 & 1 & 2 & $-$0.416 & $-$0.374\\
                     & 1 & 0 & 2 & +0.416   & +0.374\\
                     & 0 & 0 & 0 & +0.057   & +0.081\\
                     & 1 & 1 & 4 & $-$0.073 & $-$0.040\\
                     & 0 & 2 & 4 & +0.040   & +0.022\\
                     & 2 & 0 & 4 & +0.040   & +0.022\\
                     & 1 & 2 & 6 & +0.016   & +0.010\\
                     & 2 & 1 & 6 & $-$0.016 & $-$0.010\\
$^3P_1 ; L = 1$ & 0 & 0 & 2 & +0.507   & $-$0.561\\
                     & 0 & 1 & 4 & +0.025   & $-$0.006\\
                     & 1 & 0 & 4 & $-$0.025 & +0.006\\
$^1D_2 ; L = 2$ & 0 & 0 & 4 & +0.016   & +0.008\\
\hline
& $n$ & $N$ & $\rho$ & $2^+_1$   & $2^+_2$\\
$^1S_0 ; L = 2$ & 0 & 0 & 2 & +0.489   & +0.256\\
                     & 1 & 0 & 4 & +0.017   & +0.007\\
                     & 0 & 1 & 4 & $-$0.011 & $-$0.005\\
$^3P_1 ; L = 1$ & 0 & 0 & 2 & $-$0.177 & +0.338\\
$^3P_2 ; L = 1$ & 0 & 0 & 2 & $-$0.307 & +0.586\\
$^1D_2 ; L = 0$ & 0 & 0 & 2 & $-$0.489 & $-$0.256\\
                     & 0 & 1 & 4 & +0.017 & +0.007\\
                     & 1 & 0 & 4 & $-$0.011 & $-$0.005\\
\hline
& $n$ & $N$ & $\rho$ & $1^+$ & \\
$^3P_0 ; L = 1$ & 0 & 0 & 2 & +0.444 &\\
$^3P_1 ; L = 1$ & 0 & 0 & 2 & +0.384 &\\
$^3P_2 ; L = 1$ & 0 & 0 & 2 & $-$0.496 &\\
\hline
\end{tabular}
 
\caption[]{
Two-proton removal amplitudes from $^{16}$O for
states of $^{14}$C that are expected to be strongly populated in
the $^{16}$O(e,e$'$pp) reaction. These are based on the Dressed RPA
calculations described in Ref.~\cite{GeAl2},
within a model space of the $1s$ up to the $2p1f$ shells and with the
G-matrix derived from the Bonn-C potential as an effective
interaction. The quantum number $\rho$ is the total number of harmonic
oscillator quanta of the pair: $\rho = 2n+l+2N+L$ (lower case for relative
and upper case for center of mass motion).
For instance $\rho = 4$ indicates contributions from the $sd$ shell and
$\rho = 6$ from the $pf$ shell.
The energies of the listed states are largely known from experiments: 
$0^+_1$ represents the ground state of $^{14}$C, $2^+_1$ represents the sum
of the $2^+$ states at 7.01 and 8.32 MeV\cite{Ajz1315}, the $1^+$ is 
known\cite{Ajz1315} at 10.3 MeV. The $2^+_2$ was identified with a bump 
around 16 MeV observed in Ref.~\cite{OnAl}. The location of the $0^+_2$
is less clear; the strength may be fragmented over several final states
in the range between 12 and 14 MeV\cite{OnAl}.
\label{tab:Amplitudes}
}
\end{table}

\end{document}